# Surface ferrimagnetic order in RuO$_2$ film

Jiahua Lu[1,2,+], Huangzhaoxiang Chen[3,+], Zhe Zhang[1,4,+], Xinyue Wang[1,2], Donghang Xie[1,2], Bo Liu[1,2], Liang He[1,2], Yao Li[1,2], Jun Du[1,5], Zhi Wang[3,*], Junwei Luo[3], Rong Zhang[2], Yongbing Xu[1,2,4,6*], Xuezhong Ruan[1,2*]

[1]State Key Laboratory of Spintronics, Nanjing University, Suzhou 215163, China

[2]Jiangsu Provincial Key Laboratory of Advanced Photonic and Electronic Materials, School of Electronic Science and Engineering, Nanjing University, Nanjing, 210093, China.

[3]State Key Laboratory of Semiconductor Physics and Chip Technologies, Institute of Semiconductors, Chinese Academy of Sciences, Beijing 100083, China

[4]School of Integrated Circuits, Nanjing University, Suzhou 215163, China.

[5]School of Physics, Nanjing University, Nanjing, 210093, China.

[6]York-Nanjing International Joint Center in Spintronics, Department of Electronics and Physics, University of York, York, YO10 5DD, United Kingdom.

[+] These authors contributed equally to this work.

[*]Correspondence and requests for materials should be addressed to Zhi Wang (email: wangzhi@semi.ac.cn), Yongbing Xu (email: ybxu@nju.edu.cn), or to Xuezhong Ruan (email: xzruan@nju.edu.cn)

**Abstract**

RuO$_2$, widely proposed as a prototypical altermagnet, remains intensely debated with regard to its magnetic nature. Here, we demonstrate that RuO$_2$ is non-magnetic in the bulk, but possesses a spontaneous surface ferrimagnetic order. Using spin- and angle-resolved photoemission spectroscopy, we directly detect a narrow surface state with identical spin polarizations at opposite momenta and at the Brillouin-zone center, incompatible with the spin texture of any altermagnetic order. First-principles calculations identify the non-magnetic bulk state and reveal that the detected magnetism is confined to the fully oxygen-terminated surface, where the charge transfer from Ru to O at surface triggers a ferrimagnetic alignment between adjacent Ru sublattices with antiparallel moments of +0.48 μ$_B$ and –0.04 μ$_B$. Our findings provide a unified explanation reconciling debating reports on the magnetism of RuO$_2$, establishing surface ferrimagnetism as the origin of the observed magnetic signals, and distinguishing it unambiguously from altermagnetism.

Altermagnet [1-4] has recently emerged as a distinct form of magnetic material in which non-relativistic exchange interaction generates momentum-dependent spin splitting without net magnetization. This concept has rapidly reshaped the classification of magnetic materials and inspired intense efforts to identify experimental realizations [5-14]. Among proposed candidates, ruthenium dioxide ($RuO_2$) has been widely regarded as a prototypical platform owing to its simple rutile structure, metallic conductivity, and predicted high Néel temperature. However, despite extensive experimental and theoretical efforts, the existence and origin of magnetic order in $RuO_2$ remain unresolved, with mutually contradictory reports alternately supporting a bulk altermagnetic state [15-23] or a fundamentally non-magnetic ground state [24-29].

Experimental evidence for magnetism in $RuO_2$ has been primarily inferred from signatures of time-reversal symmetry (TRS) breaking, including anomalous Hall responses [15], spin-current generation [16], terahertz emission [17,18], X-ray magnetic circular dichroism [19], and spin splitting effect [20-23]. However, TRS breaking is essential but not sufficient for altermagnetism as it can arise from any form of local or itinerant magnetic order [30-32], as well as from extrinsic effects. Crucially, the defining hallmark of altermagnetism, i.e., the coexistence of TRS breaking with an antiferromagnetic arrangement that generates non-relativistic spin splitting in the band structure, has not been unambiguously established in $RuO_2$. In particular, spin- and angle-resolved photoemission experiments, which directly probe momentum-resolved spin textures, have so far failed to detect the characteristic altermagnetic band splitting expected for a bulk altermagnet [24,25]. Meanwhile, a growing body of bulk-sensitive probes points toward a non-magnetic ground state of $RuO_2$. Muon spin rotation

[27,28], quantum oscillation measurements [29], and optical conductivity [30] consistently report electronic and magnetic properties that are well described by non-magnetic calculations, placing stringent upper bounds on any possible bulk magnetic moment. The coexistence of apparent TRS-breaking signals with the absence of detectable bulk magnetism has thus created a fundamental inconsistency that cannot be resolved within a purely bulk-based picture of altermagnetism or non-magnetism.

Here we address this critical controversy by demonstrating that $RuO_2$ is intrinsically non-magnetic in the bulk, while exhibiting a spontaneous ferrimagnetic order confined to an oxygen-rich surface termination. By combining spin- and angle-resolved photoemission spectroscopy (spin-ARPES) with first-principles calculations on high-quality (110)-oriented $RuO_2$ film, we observe a narrow spin-polarized surface band whose polarization persists at opposite momenta and at the Brillouin-zone center, a behavior fundamentally incompatible with any bulk altermagnetic spin texture. Our first-principles calculations reveal that the bulk state is intrinsically non-magnetic, but the enhanced density of states (DOS) at the fully oxygen-terminated surface due to charge transfer from Ru to O drives a Stoner instability. Our study provides a coherent framework that reconciles previous debates on $RuO_2$ magnetism and establishes surface-induced ferrimagnetism, rather than altermagnetism, as the origin of the observed magnetic signals.

A 15 nm-thick single-crystal $RuO_2$ film was deposited onto (110)-oriented $TiO_2$ substrate with magnetron sputtering system (see details in supplement information (SI) [33]) for spin-ARPES measurements. The reflection high energy electron diffraction (RHEED) patterns, as shown in Fig. 1(a) and (b), indicate high crystalline quality with a flat well-ordered surface.

The momentum-space cut $\overline{\Gamma} - \overline{M}$, marked by the red line in Fig. 1(d), is identified via the low energy electron diffraction (LEED) pattern in Fig. 1(c). The valence band (VB) dispersion along $\overline{\Gamma} - \overline{M}$ direction after the implementation of the second derivative with respect to energy is displayed in Fig .1(e). Two narrow bands are situated near 0.1 eV (α) and 0.5 eV (β) below the Fermi level with a parabolic dispersive band (δ) passing through them, as marked by the red dashed lines in Fig. 1(e).

The calculated spectral intensity based on surface slab models (see SI for details) [33] reproduces all the essential characteristics of the observed α, β and δ bands, as shown in Fig. 1(f). Four distinct surface terminations are considered, labeled $S_{4-5}$, $S_{6-5}$, $S_{4-6}$ and $S_{6-6}$, according to coordination number of the two adjacent but inequivalent Ru sublattices at surface (Fig. S1) [33]. The practical surface configuration can be modelled by mixing the four surface terminations with relative proportions determined by the formation energies and temperature. As summarized in Table I, the half-oxygen-covered surface ($S_{6-5}$) is the most stable configuration under both oxygen-poor (O-poor) and oxygen-rich (O-rich) circumstances. In an O-poor atmosphere, all other surface terminations vanish completely, while in an O-rich atmosphere, the fully O-covered surface ($S_{6-6}$) becomes nearly degenerate in energy with $S_{6-5}$, with a formation energy of only ~0.08 eV/atom higher than that of $S_{6-5}$. As a consequence, under the growth temperature of 773 K and an O-rich atmosphere, the surface should consist of approximately 77% $S_{6-5}$ and 23% $S_{6-6}$ configurations. The experimental band structure in Fig. 1(e) is well captured under such composition. The β and δ bands arise entirely from $S_{6-5}$, while the α band is predominantly derived from $S_{6-5}$ with minor contribution from $S_{6-6}$.

To explore the origin of magnetism, we extract the energy distribution curves (EDCs) at the $\bar{\Gamma}$ point (C$_2$) and at the symmetric momenta ±k (C$_1$, C$_3$) along the $\bar{\Gamma}-\bar{M}$ direction, as marked by the yellow dashed lines in Fig. 1(e). The second derivatives of these EDCs are shown in Fig. 2(a)-(c), respectively. All the EDCs exhibit consistent peak positions at the α and β bands, while the δ band reaches the minimum energy at C$_2$, all consistent with the band dispersion in Fig. 1(e). Fig. 2(d)-(f) present the corresponding spin polarizations. A pronounced spin polarization of approximately 20% is detected at the α band, while the β and δ bands exhibit no measurable spin polarization within experimental resolution. Crucially, the spin polarizations at +k (C$_3$) and −k (C$_1$) possess the same orientation along the $\bar{\Gamma}-\bar{M}$ direction and remain finite at $\bar{\Gamma}$ point (C$_2$). This behavior directly contradicts the symmetry requirement of bulk altermagnetism in RuO$_2$ which predicts opposite spin polarization at ±k and zero spin polarization at $\bar{\Gamma}$ point [38]. As presented in Fig. S2 [33], the energy dispersion along $\bar{\Gamma}-\bar{Z}$ direction also exhibits the spin-polarized narrow band at ~0.1 eV below the Fermi level, which also disagrees with the altermagnetism-based calculation in which such a spin polarized narrow band only appears along $\bar{\Gamma}-\bar{M}$ direction [38]. Therefore, the altermagnetism cannot be the origin of the observed spin polarization, nor can the presence of such narrow bands be regarded as a spectral signature of altermagnetism in RuO$_2$. The identical spin orientation observed at opposite momenta also rules out a Rashba-like spin-orbit coupling origin, as in spin-orbit-driven surface states the spin orientation is locked to momentum and must reverse under k → −k. The total magnetic moment of pure 15 nm-thick RuO$_2$ obtained by vibrating sample magnetometer is around 5×10$^{-7}$ emu (Fig. S3(b)) [33]. If normalized by the full film thickness of 15 nm, the saturation magnetization is on the order of 10 emu/cm$^3$. Such value is roughly two orders of magnitude smaller than that of a classic ferromagnet and thus rules out the presence of bulk ferromagnetism in RuO$_2$ film. When normalized by only the topmost two atomic layers, the estimated saturation magnetization is ~389 emu/cm$^3$, indicating that the observed spin polarization originates predominantly from the surface region.

Previous first-principles calculations based on a bulk-altermagnetism ground state [19,38] predict dispersive bands with strongly momentum-dependent spin splitting. Such features are obviously inconsistent with either the narrow α and β bands or the spin texture. Here, our theoretical model reproduces both the magnitude (~16%) and the energy position of the spin polarization peak observed experimentally, as indicated within the energy interval marked by the purple dashed lines in Fig. 3(a) and Fig. 3(b). In addition, the calculation predicts a secondary spin-polarized band (γ) approximately 1 eV below the Fermi level with opposite spin orientation. Experimentally, this feature appears only as a weak shoulder in the second derivative of EDC, as marked by the blue dashed line in Fig. 2(b). The subtle shoulder peak indicates that the intensity of secondary-electron background is comparable to that of γ band and the spin polarization of γ band is obscured by the unpolarized secondary-electron background. More precisely, our spin-polarized calculations reveal a decisive contrast among the bulk and the four candidate surface terminations. Bulk RuO$_2$, as well as the most stable surface configuration S$_{6-5}$, possess non-magnetic ground states. Only the fully oxygen-terminated S$_{6-6}$ stabilizes a ferrimagnetic alignment between two inequivalent Ru sites. In fact, the energy lowering from non-magnetic to magnetic states on S$_{6-6}$ is approximately 19 meV per formula unit on the surface, or 3.1 μJ/cm$^2$, indicating that the non-magnetic S$_{6-6}$ is energetically unstable. The Ru$_1$ site bonded to two surface O carries a larger magnetic moment of +0.48 $\mu_B$, while the Ru$_2$ site bonded to single surface O hosts a smaller antiparallel magnetic moment of

-0.04 $\mu_B$. This ferrimagnetic pattern is confined to the topmost two layers, while the deeper layers remain non-magnetic, as shown in Fig. 3(c).

The calculated surface state under O-poor circumstance, dominated by $S_{6-5}$, is purely spin-unpolarized and consistent with the experimental results of (110)-oriented $RuO_2$ film grown by molecular beam epitaxy (MBE) [25]. In contrast, the calculated surface state under O-rich circumstance incorporates approximately 23% spin-polarized $S_{6-6}$ termination, which is in agreement with the experimental results of our (110)-oriented $RuO_2$ film grown by magnetron sputtering system. Unlike the high-vacuum environment of MBE, the continuous introduction of $O_2$ during the deposition with magnetron sputtering promotes the formation of the $S_{6-6}$ termination, and hence the ferrimagnetic order. Fig. 3(d)-(g) presents the calculated energy dispersion of spin-up and spin-down channels projected and normalized on the surface $S_{6-5}$ and $S_{6-6}$, respectively. Although dominated by the $S_{6-5}$ termination, the narrow band α in the two spin channels from $S_{6-5}$ termination are degenerate and show no spin polarization. The spin polarization of α band originates solely from the $S_{6-6}$ termination. This explains the relatively low but non-negligible spin polarization of the band α. The narrow γ band is completely vanishing in $S_{6-5}$, and hence exclusively associated with the $S_{6-6}$ termination. Owing to the low surface coverage, the DOS of γ band is not strong enough for precise detection, consistent with the experimental observations noted above.

We note that several theoretical works have attributed surface magnetism in (110)-oriented rutile structure to the symmetry breaking of coordination environment (SBCE) for distinct sublattice atoms at the surface [39,40]. Within this framework, the partially oxygen-terminated $S_{4-5}$, $S_{4-6}$, and $S_{6-5}$ [41] configurations are expected to be magnetic. In contrast, our results show that SBCE instead quenches magnetic orders. The physical origin of this surface magnetism lies in the charge redistribution between surface Ru and O. Increasing the oxygen coverage transfers electrons from Ru 4d orbitals to surface O 2p orbitals, resulting in a depletion of occupation on Ru sites (Table I). This redistribution shifts the narrow flat band within the valence manifold toward the Fermi level on surface, and substantially enhances the local DOS $D(E_F)$. Within the Stoner picture, spontaneous spin polarization occurs when the exchange energy gain exceeds the kinetic energy cost, expressed by the criterion $ID(E_F) > 1$ [41-43]. The Ru sites on the fully O-terminated $S_{6-6}$ surface lose more charge (~0.1 e$^-$ per Ru) and experience a larger shift in $E_F$ than that of other surfaces. Reported effective Stoner parameters for Ru 4d states in metallic ruthenates lie in the range $I \approx 0.4$-$0.6$ eV, depending on compound and computational definition [44,45]. Within such a range, we find that $ID(E_F)$ exceeds unity only for $Ru_1$ site at $S_{6-6}$ configurations while staying well below unity for bulk and all other surface configurations (Fig. S4) [33]. Therefore, a Stoner instability is triggered only on the fully O-terminated $S_{6-6}$ surface, while all other configurations remain non-magnetic. Given that density-functional theory often overestimates the bandwidth and underestimate the effective mass in transition-metal oxides [46-48], such close agreement represents strong validation of our model of the surface ferrimagnetic order.

In summary, our study provides a coherent framework for understanding the magnetic signatures reported in $RuO_2$, and addresses the long-standing debate over whether it hosts a bulk altermagnetic state or is fundamentally non-magnetic. We demonstrate that the magnetic signals detected in $RuO_2$ violate the symmetry constraints required for bulk altermagnetism and is therefore incompatible with an intrinsic altermagnetic ground state. Such magnetism originates not from the bulk state, but from a previously overlooked surface state. Through direct spin-resolved photoemission spectroscopy on high quality RuO$_2$ film and first-principles calculations, we identify a spontaneous ferrimagnetic order confined exclusively to the fully oxygen terminated surface. This surface magnetism arises from charge redistribution rather than the intrinsic spin splitting expected for altermagnetism. Our work not only clarifies the origin of contradictory reports in RuO$_2$ but also establishes surface induced magnetism as a crucial and general mechanism that can be manifested in other non-magnetic materials, with significant implications for the interpretation of any magnetic-related signals and the design of interfacial spintronic systems.

**Acknowledgments**

The authors thank Prof. Shan Qiao for fruitful discussions. This work is funded by the National Key Research and Development Program of China (No. 2024YFA1408801), National Natural Science Foundation of China (Grant No. 62427901, T2394473), and the Natural Science Foundation of Jiangsu Province of China (BK20211144).

**Author Contributions:**

X. Ruan, Y. Xu, Z. Wang, and J. Lu conceived the project; J. Lu grew the $RuO_2$ sample with the help of Z. Zhang; J. Lu performed the ARPES experiments with the help of L. He, X. Ruan, B. Liu, and Y. Li; J. Lu performed the VSM measurements with the help of J. Du and D. Xie; J. Lu analyzed all the experimental data with the aid of X. Wang and X. Ruan; Z. Wang and H. Chen developed and performed the calculations and analyzed the calculated data; J. Lu, Z. Wang and H. Chen wrote the manuscript with contributions from all authors

**Competing interests declaration**: The authors declare no conflicts of interests.

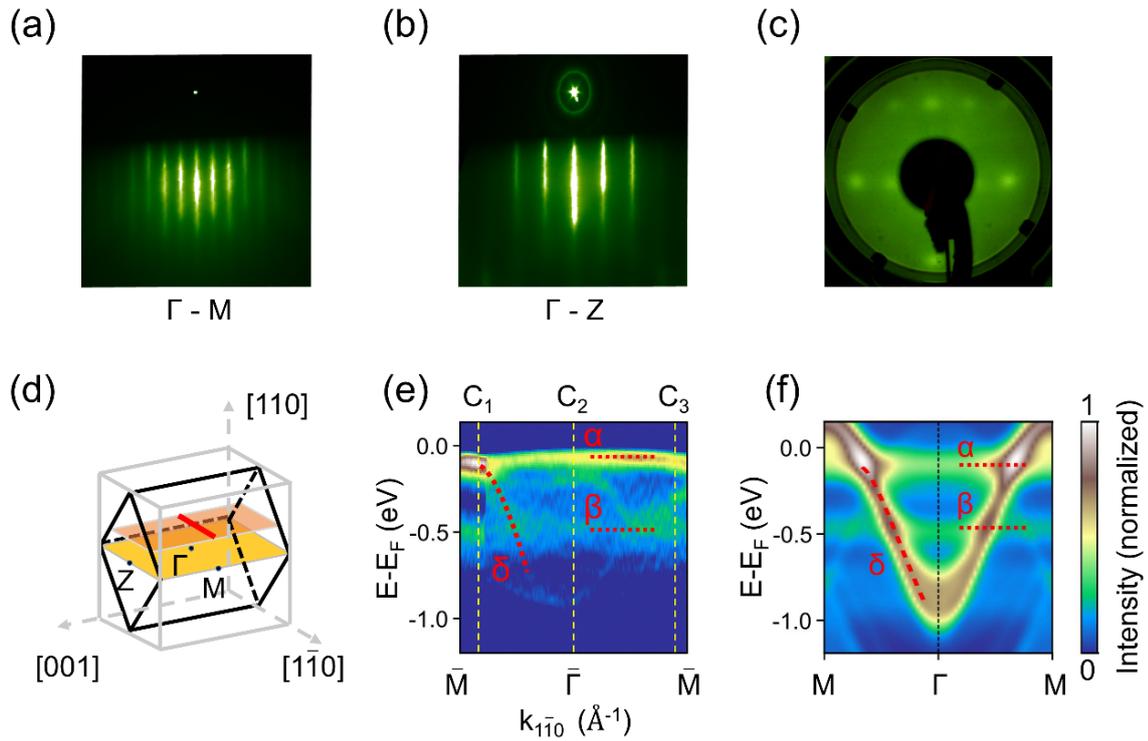

FIG. 1 RHEED patterns along Γ – M (a) and Γ – Z (b) directions. (c) LEED pattern for identifying the $\bar{\Gamma} - \bar{M}$ direction. (d) Schematic of the three-dimensional Brillouin zone of $RuO_2$. The $\bar{\Gamma} - \bar{M}$ direction is marked by the red line. (e) The second derivative of the band structure along $\bar{\Gamma} - \bar{M}$ direction. The yellow dashed lines $C_1$, $C_2$, $C_3$ are the locations of EDCs for spin polarization measurements. The bands α, β, and δ are labeled by the red dashed lines. (f) Energy dispersion obtained by the surface model calculations centered at Γ point along Γ - M direction.

TABLE I. Surface formation energies and site-resolved electronic properties of (110)-oriented $RuO_2$ terminations. Calculated formation energies under O-rich and O-poor limits, Ru 4d charge variations relative to bulk, and magnetic moments for the four candidate surface terminations ($S_{4-5}$, $S_{6-5}$, $S_{4-6}$, and $S_{6-6}$).

| Surface configuration | Formation energy (eV/atom) with respect to $S_{6-5}$ | | Surface Ru local charge ($e^-$) with respect to bulk | | Surface Ru local moment ($\mu_B$) |
|---|---|---|---|---|---|
| | O-rich | O-poor | | | |
| $S_{4-5}$ | +2.47 | +0.92 | $Ru_1$ | 0 | 0 |
| | | | $Ru_2$ | -0.02 | 0 |
| $S_{6-5}$ | 0 | 0 | $Ru_1$ | -0.05 | 0 |
| | | | $Ru_2$ | -0.05 | 0 |
| $S_{4-6}$ | +2.10 | +2.02 | $Ru_1$ | -0.06 | 0 |
| | | | $Ru_2$ | -0.06 | 0 |
| $S_{6-6}$ | +0.08 | +1.57 | $Ru_1$ | -0.09 | 0.48 |
| | | | $Ru_2$ | -0.08 | -0.04 |

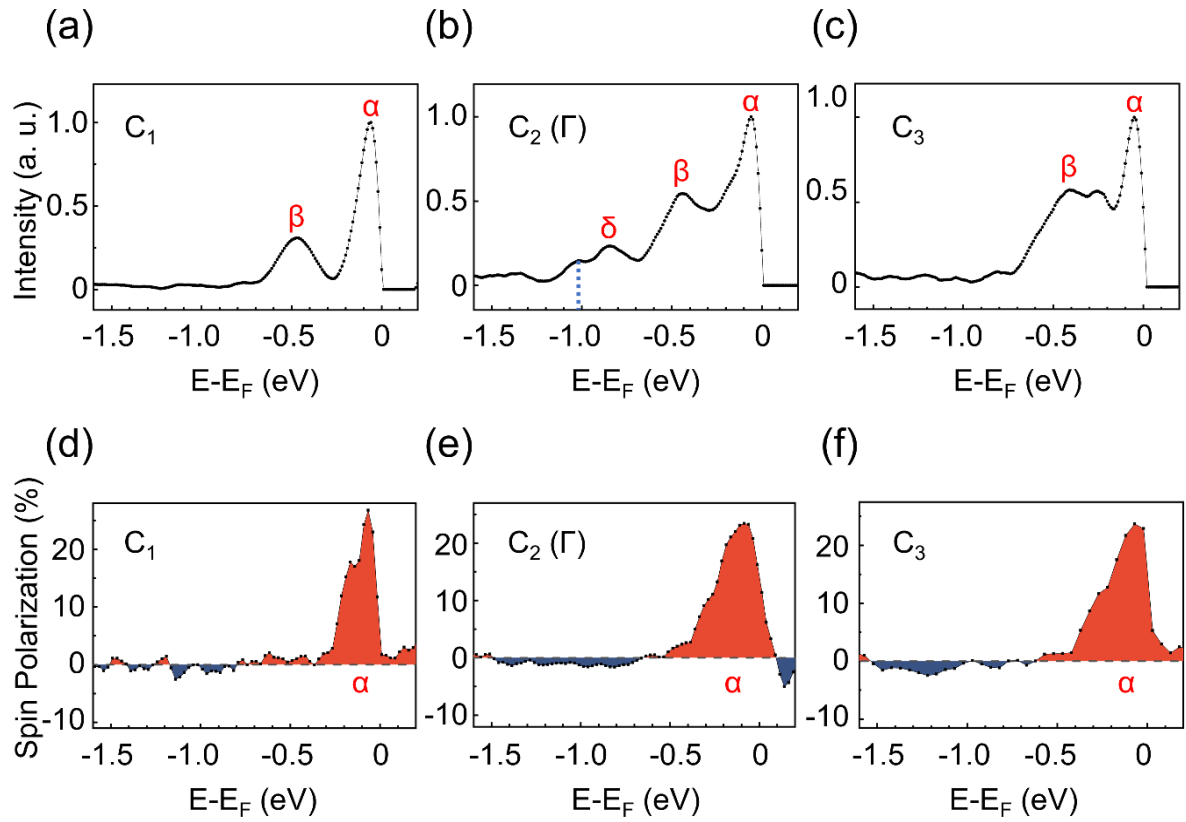

FIG. 2 Second derivatives of the EDCs at $C_1$ (a), $C_2$ (b), and $C_3$ (c). The pronounced peaks of α, β and δ bands are labeled with the red letters. Corresponding spin-polarizations spectra for $C_1$ (d), $C_2$ (e), and $C_3$ (f), respectively.

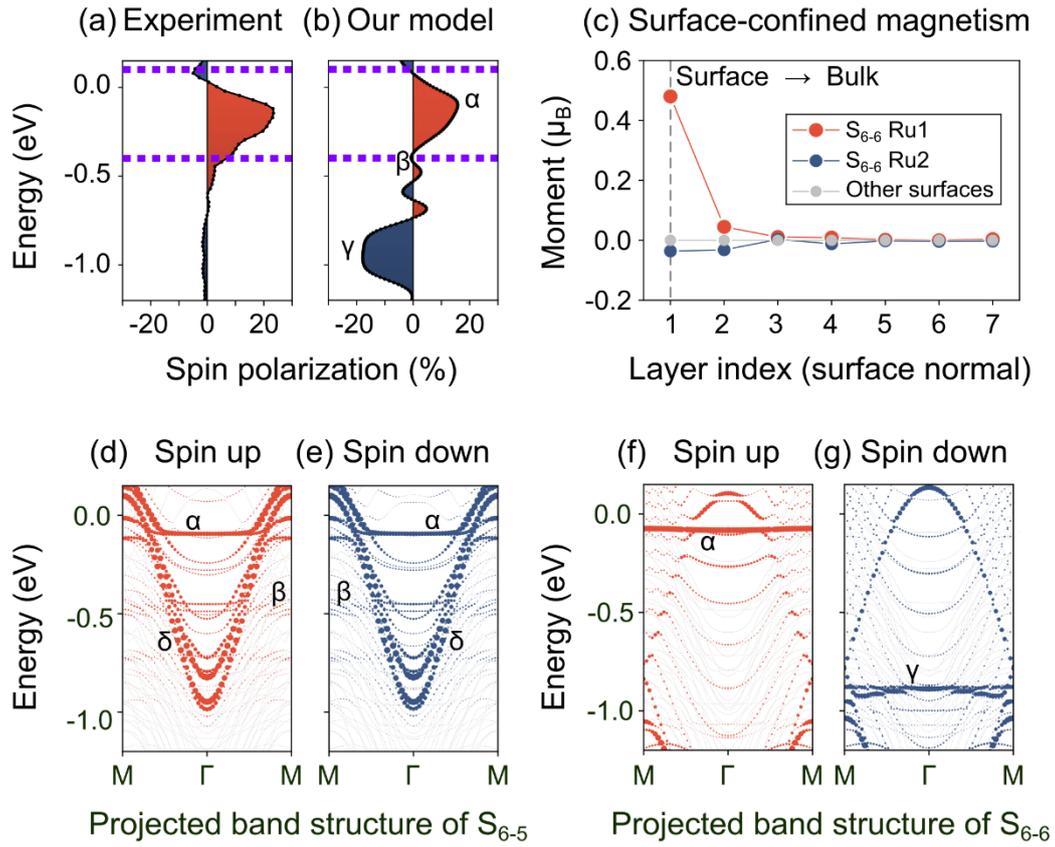

FIG. 3 (a) Spin polarization of EDC at $\bar{\Gamma}$ point (C$_2$), showing a pronounced polarization at the α band. (b) Calculated spin polarization of EDC at Γ point from the surface model. (c) Layer-resolved magnetic moments of Ru atoms for different surface terminations. Finite moments are confined to the outermost layers of the fully oxygen-terminated S$_{6-6}$ surface (The red and blue denote opposite spin orientations), whereas S$_{4-5}$, S$_{4-6}$ and S$_{6-5}$ remain non-magnetic (grey). Spin-resolved band structures projected onto S$_{6-5}$ (d,e) and S$_{6-6}$ (f,g) surface terminations. Light-gray bands indicate the projected bulk states. Spin splitting appears exclusively for S$_{6-6}$, identifying the fully oxygen-terminated surface as the origin of the observed momentum-resolved spin polarization.

# Supplement information for
# Surface ferrimagnetic order in RuO$_2$ film

Jiahua Lu[1,2,+], Huangzhaoxiang Chen[3,+], Zhe Zhang[1,4,+], Xinyue Wang[1,2], Donghang Xie[1,2], Bo Liu[1,2], Liang He[1,2], Yao Li[1,2], Jun Du[1,5], Zhi Wang[3,*], Junwei Luo[3], Rong Zhang[2], Yongbing Xu[1,2,4,6,*], Xuezhong Ruan[1,2,*]

[1]State Key Laboratory of Spintronics, Nanjing University, Suzhou 215163, China

[2]Jiangsu Provincial Key Laboratory of Advanced Photonic and Electronic Materials, School of Electronic Science and Engineering, Nanjing University, Nanjing, 210093, China.

[3]State Key Laboratory of Semiconductor Physics and Chip Technologies, Institute of Semiconductors, Chinese Academy of Sciences, Beijing 100083, China

[4]School of Integrated Circuits, Nanjing University, Suzhou 215163, China.

[5]School of Physics, Nanjing University, Nanjing, 210093, China.

[6]York-Nanjing International Joint Center in Spintronics, Department of Electronics and Physics, University of York, York, YO10 5DD, United Kingdom.

[+] These authors contributed equally to this work.

[*]Correspondence and requests for materials should be addressed to Zhi Wang (email: wangzhi@semi.ac.cn), Yongbing Xu (email: ybxu@nju.edu.cn), or to Xuezhong Ruan (email: xzruan@nju.edu.cn)

**Synthesis of RuO₂**

The 15 nm thick RuO₂ film was grown on (110)-oriented TiO₂ substrate by magnetron sputtering system. A pure ruthenium target was used with the radio frequency (RF) power set at 50 W. The substrate temperature was fixed at 773 K during the deposition. The Ar gas flow was controlled at 40 standard cubic centimeter per minute (sccm), while the O₂ flow rate was controlled at 10 sccm. The film was annealed at 773 K for 10 minutes in 10 sccm O₂ after deposition. The RHEED analysis confirmed that the film was fully relaxed with lattice parameters of a=b=4.48 Å and c=3.11 Å, compared to the RHEED pattern of TiO₂ with lattice parameters of a=b=4.6 Å and c=2.95 Å.

**ARPES and spin-ARPES measurements**

A helium lamp was used to generate ultraviolet photons with an energy of 21.218 eV (He-I). The base pressure of the main chamber of ARPES system was $1 \times 10^{-10}$ mbar. The grown RuO₂ sample was directly transferred to the main chamber via a tube system with pressure of around $2 \times 10^{-9}$ mbar. Data were measured with a SPECS PHOIBOS 150 hemispherical energy analyzer. The angular resolution of the system is 0.05° and the energy resolution is 35 meV at room temperature. Spin-ARPES measurements were realized via a one dimension in-plane Micro-Mott spin detector that includes a strong spin-orbital coupling thorium target and a couple of electron multipliers to collect the scattered electrons. The spin polarization $SP(\varepsilon)$ at energy $\varepsilon$ was obtained by

$$SP(\varepsilon) = \frac{1}{S}\frac{I_\uparrow(\varepsilon) - I_\downarrow(\varepsilon)}{I_\uparrow(\varepsilon) + I_\downarrow(\varepsilon)} \tag{1}$$

where Sherman function of the equipment is $S = 0.16$ and $I_{\uparrow/\downarrow}(\varepsilon)$ are the intensity collected by a couple of electron multipliers subtracted by the same Shirley background.

A low energy electron diffraction (LEED) was attached to the main chamber to identify the momentum cut direction. The entrance slit of the hemispherical analyzer was aligned horizontally relative to the sample, hence the horizontal direction of LEED corresponded to the momentum direction measured by ARPES.

**First-principles calculations**

Electronic-structure calculations were performed using the Vienna Ab initio Simulation Package (VASP) [1] within the projector-augmented wave (PAW) formalism [2]. The Perdew–Burke–Ernzerhof (PBE) functional of the generalized gradient approximation (GGA) was adopted [3], with an energy cutoff of 560 eV. Brillouin-zone sampling employed Monkhorst-Pack meshes of 8 × 8 × 1 for slab calculations and 8 × 8 × 8 for bulk calculations. Unless otherwise specified, the experimental lattice parameters (a = 4.48 Å, c = 3.11 Å) were employed without further relaxation to maintain consistency with ARPES geometry. To assess correlation effects on Ru 4d states, we performed additional DFT + U [4,5] tests with U varied from 0 to 2 eV. The results reported in the main text were obtained at U=0, while the emergence of surface magnetism on the fully oxygen-terminated surface and the non-magnetic character of the other terminations remain robust across U= 0-2 eV.

**Surface terminations and formation energies**

Four (110)-oriented $RuO_2$ surface terminations with increasing oxygen coverage were considered [6,7] (Fig. S1): $S_{4-5}$ (fully Ru-terminated), $S_{6-5}$ (half-O-covered, where all Ru$_1$ atoms are passivated by O atoms while Ru$_2$ atoms remain bare), S$_{4-6}$ (half-O-covered, where Ru$_2$ are O-covered while Ru$_1$ are bare), and S$_{6-6}$ (fully oxygen-terminated). Symmetric slab models containing up to 13 atomic layers were constructed, separated by 15 Å thick vacuum. Dipole corrections were applied along the surface normal to remove spurious electrostatic fields. The surface formation energies were computed as

$$E_f(i) = \left(E_{slab}^i - E_{slab}^1 + n_O(i)\mu_O\right)/2 \qquad (2)$$

where $E_{slab}^{(i)}$ is the total energy of the slab with termination $i$, $E_{slab}^{(1)}$ is the reference slab energy (here, S$_{6-5}$), and $n_O(i)$ denotes the difference in the number of oxygen atoms between termination and the reference. The oxygen chemical potential $\mu_O$ was varied to represent oxygen-rich and oxygen-poor limits. For each termination, spin-polarized self-consistent calculations were performed to obtain the layer-resolved magnetic moments and charge distributions.

**EDC simulation**

The theoretical EDCs at Γ point were calculated by summing the projected density of states (PDOS) of atoms within one bulk lattice constant from the surface for each termination, weighted by the equilibrium surface fractions obtained from the formation-energy analysis. Photoemission matrix-element effects were neglected for clarity. The spin polarization $P(\varepsilon)$ at energy $\varepsilon$ was computed as

$$P(\varepsilon) = \frac{\text{PDOS}_\uparrow(\varepsilon) - \text{PDOS}_\downarrow(\varepsilon)}{\text{PDOS}_\uparrow(\varepsilon) + \text{PDOS}_\downarrow(\varepsilon)} \qquad (3)$$

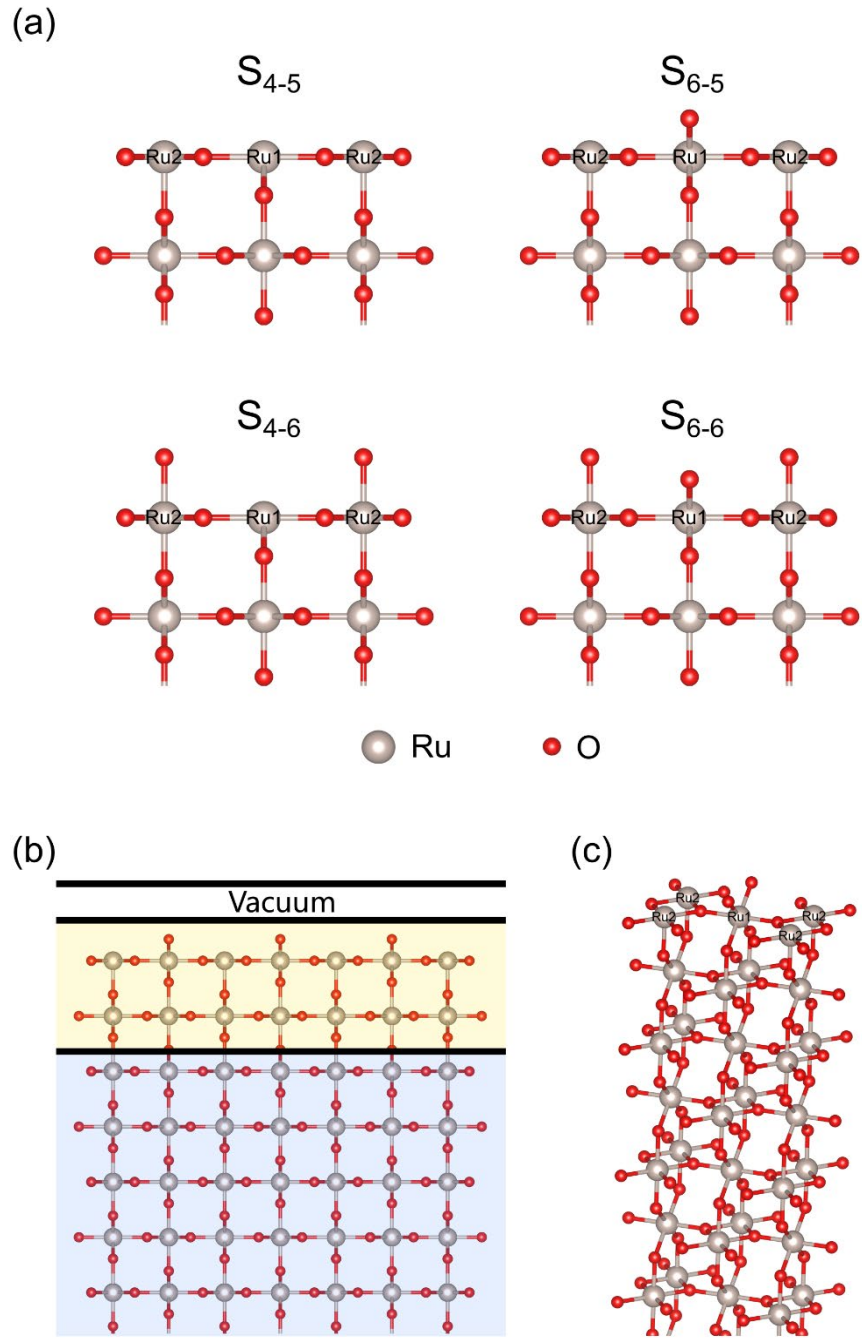

FIG. S1 (a) Four kinds of $RuO_2$ surface structures. (b) The overall structure consisting of magnetic surface (yellow) and non-magnetic bulk (blue). (c) Geometrical structure of $RuO_2$(110) surface.

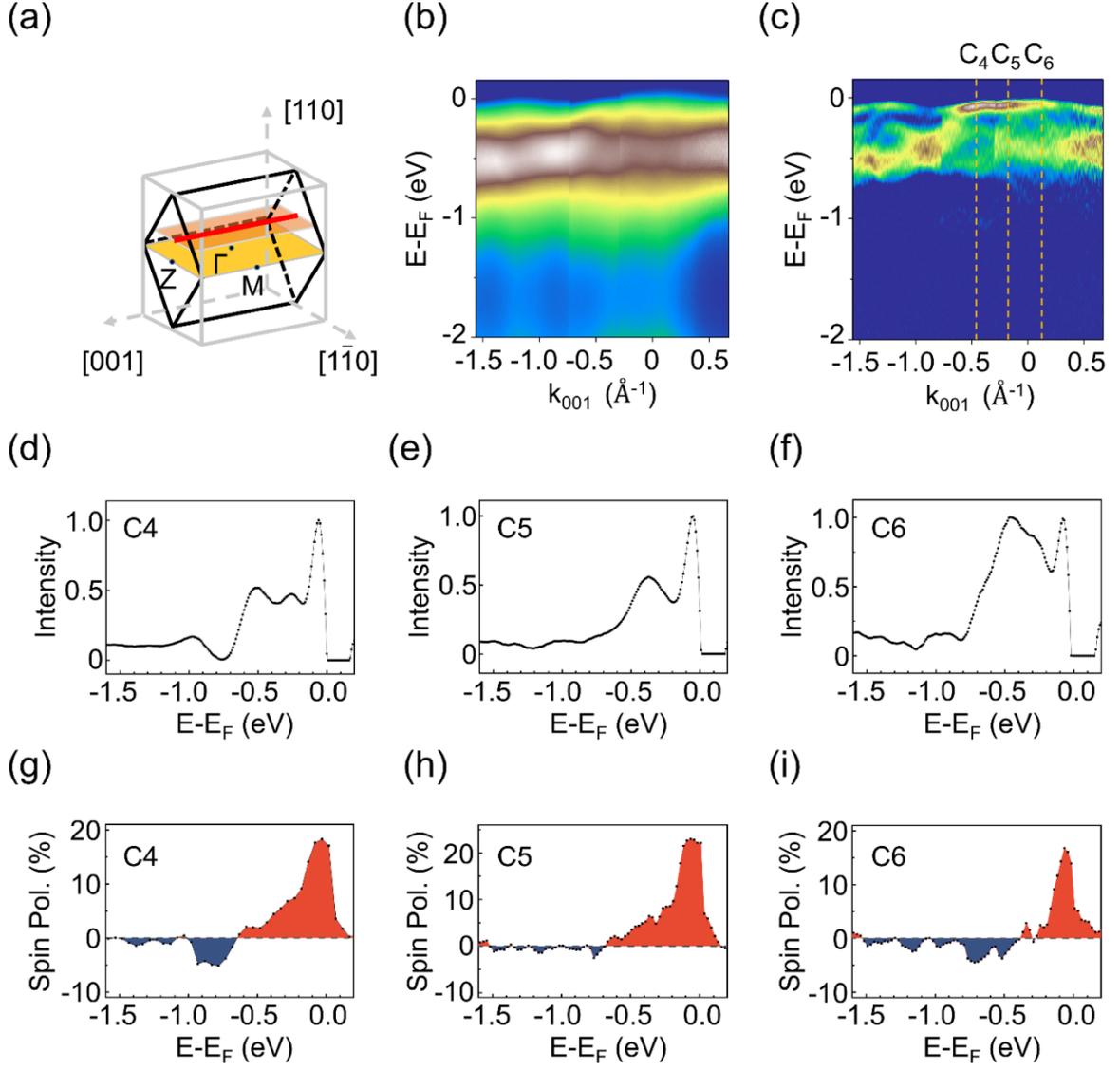

FIG. S2 (a) Schematic of the three-dimensional Brillouin zone of RuO$_2$. The $\bar{\Gamma} - \bar{Z}$ direction is marked by the red line. (b) Measured electronic band along $\bar{\Gamma} - \bar{Z}$ direction. (c) Second derivatives of the band structure in (b). The yellow dashed lines C$_4$, C$_5$, C$_6$ are the locations of EDCs for spin-ARPES measurements. Second derivatives of C$_4$ (d), C$_5$ (e), C$_6$ (f) and their corresponding spin polarizations (g, h, i), respectively.

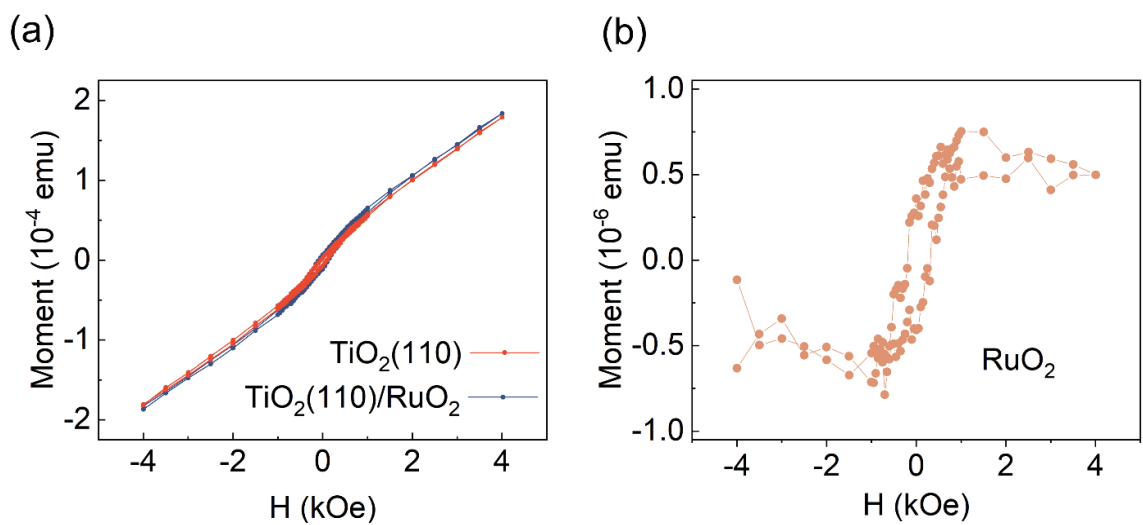

FIG. S3 (a) Magnetic hysteresis loops measured at room temperature for the bare (110)-oriented $TiO_2$ substrate (red) and the 15 nm-thick $RuO_2$ (black) film grown on the same substrate. (b) Magnetic hysteresis loop of the 15 nm-thick $RuO_2$ film, which is obtained through subtracting the magnetic moment of the bare $TiO_2$ substrate from that of $TiO_2(110)/RuO_2$.

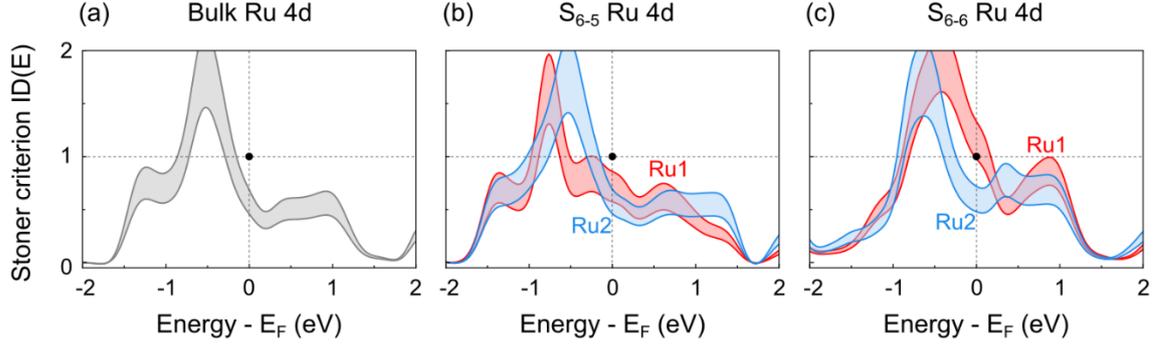

FIG. S4 Energy-dependent Stoner product $ID(E)$ in RuO$_2$. (a) Bulk. (b) S$_{6\text{-}5}$ surface. (c) S$_{6\text{-}6}$ surface. Shaded regions represent $I$=0.4-0.6 eV. The vertical dash lines mark $E - E_F = 0$, and the horizontal lines indicate the Stoner threshold $ID(E) = 1$. Only the fully oxygen-terminated S$_{6\text{-}6}$ surface satisfies $ID(E_F) > 1$. All $D(E)$ are evaluated using the non-spin-polarized electronic structures as the paramagnetic reference states. $D(E)$ on surfaces are extracted from the outermost two layers.